\documentclass[runningheads]{llncs}
\usepackage[T1]{fontenc}
\usepackage{graphicx}
\usepackage{enumitem}
\usepackage{listings}
\usepackage{xcolor}
\usepackage{capt-of}
\usepackage{url}        
\usepackage{breakurl}   
\usepackage{multirow}
\usepackage{amsmath}
\usepackage{tabularx}
 
\begin{document}
\title{Quantifying the Energy Consumption and Carbon Emissions of LLM Inference via Simulations}

\titlerunning{Quantifying Energy and Emissions of LLM Inference} 

\author{
Miray Özcan\inst{1} \and 
Philipp Wiesner\inst{2} \and 
Philipp Weiß\inst{2} \and 
Odej Kao\inst{2} 
}

\authorrunning{Özcan et al.}

\institute{
Minerva University, California, USA\\
\email{miray@uni.minerva.edu}
\and
Technische Universität Berlin, Germany\\
\email{\{wiesner, weiss, odej.kao\}@tu-berlin.de}
}

\maketitle
\vspace{-1.5em}
\begin{abstract}
The environmental impact of Large Language Models (LLMs) is rising significantly, with inference now accounting for more than half of their total lifecycle carbon emissions. However, existing simulation frameworks, which are increasingly used to determine efficient LLM deployments, lack any concept of power and, therefore, cannot accurately estimate inference-related emissions. We present a simulation framework to assess the energy and carbon implications of LLM inference under varying deployment setups. First, we extend a high-fidelity LLM inference simulator with a GPU power model that estimates power consumption based on utilization metrics, enabling analysis across configurations like batch size, sequence length, and model parallelism. Second, we integrate simulation outputs into an energy system co-simulation environment to quantify carbon emissions under specific grid conditions and explore the potential of carbon-aware scheduling. Through scenario-based analysis, our framework reveals how inference parameters affect energy demand and carbon footprint, demonstrates a renewable offset potential of up to 69.2\% in an illustrative deployment case, and provides a foundation for future carbon-aware inference infrastructure design.

\keywords{LLM inference \and GPU power modeling \and energy efficiency \and sustainable computing}
\end{abstract}
%
%
%

\section{Introduction}

Large language models (LLMs) are increasingly deployed across cloud data centers and HPC clusters, where inference workloads are highly concurrent and distributed. Their cumulative energy consumption during inference now rivals or exceeds training \cite{wu2022sustainableai,samsi2023from}. A single LLM query may consume just 0.3–1~Wh, but with millions of daily requests, emissions scale rapidly~\cite{faiz2024llmcarbon,wiesner2025qualitytime}. By 2026, global inference demand is projected to reach petawatt-hour levels \cite{patterson2021carbonemissions}.

Yet LLM inference remains operationally inefficient. Throughput bottlenecks from large model sizes, limited memory bandwidth, and static GPU power draw undermine energy proportionality \cite{agrawal2024vidur,fernandez2024hardware}. Existing power models are typically tailored to offline training and rely on coarse, system-level telemetry \cite{anthony2020carbontracker,courty2024codecarbon}, overlooking inference-specific inefficiencies and token-level variability.

Recent systems research has begun to address sustainability challenges in ML workloads. LLMCarbon~\cite{faiz2024llmcarbon} offers lifecycle estimates but assumes static utilization. SPROUT~\cite{li2024sprout} modifies the number of tokens generated based on grid conditions, while EcoServe~\cite{li2025ecoserve} optimizes resource provisioning over multi-hour job traces. However, none simulate token-level execution dynamics or enable sub-second emissions modeling in LLM inference pipelines.

To address these gaps, we introduce a simulation framework that integrates Vidur~\cite{agrawal2024vidur}, a high-fidelity LLM inference simulator, with Vessim~\cite{wiesner2024vessim}, a grid-aware energy co-simulator. We (1) extend Vidur with a GPU power model to estimate energy use under varying parameters (e.g., input length, batch size, parallelism), and (2) link its outputs to Vessim to evaluate emissions under dynamic grid conditions, including solar availability and carbon intensity.
This enables pre-deployment exploration of energy–performance trade-offs across hardware setups and scheduling policies without physical infrastructure or real-world emissions.
All code is publicly available at \url{https://github.com/ozcanmiraay/vidur-energy}.

\section{The Need for Energy Tracking in LLM Inference}

Despite the rising footprint of inference, no existing tools provide fine-grained energy or emissions modeling tailored to LLM deployments. Tools like CarbonTracker~\cite{anthony2020carbontracker} and CodeCarbon~\cite{courty2024codecarbon} offer coarse estimates based on average GPU draw but are designed for training.
Simulators such as Vidur~\cite{agrawal2024vidur} enable detailed exploration of latency and throughput trade-offs across batch sizes, parallelism strategies, and scheduling policies. Vidur integrates with production inference engines like vLLM~\cite{kwon2023pagedattention} and logs token-level performance metrics, but lacks any notion of energy or carbon impact.

This reveals a key gap: energy tracking tools lack LLM-specific resolution, while simulators lack emissions modeling. We address this intersection by integrating a GPU power model into Vidur and connecting it to Vessim~\cite{wiesner2024vessim} to support sub-second emissions simulation. This enables carbon-aware design exploration in realistic grid conditions without requiring physical deployments.
A central challenge lies in quantifying GPU workload intensity. Common tools like NVML report utilization based on kernel occupancy but fail to differentiate between compute-bound and memory-stalled execution~\cite{fernandez2024hardware}. This is especially misleading during decoding, where utilization remains high but energy efficiency drops~\cite{agrawal2024taming}.
To address this, we adopt \emph{Model FLOPs Utilization (MFU)}~\cite{chowdhery2023palm}:
\[
\text{MFU} = \frac{\text{Achieved FLOPs/sec}}{\text{Peak FLOPs/sec}}.
\]
MFU better reflects arithmetic saturation and correlates with dynamic power. For example, when MFU drops by 30\%, power may decline by under 10\%~\cite{fernandez2024hardware}, highlighting static draw and inefficiency in decode-heavy stages.

While Vidur outputs MFU traces, it does not estimate power or carbon emissions. Our extension addresses this gap by enabling energy and emissions modeling for novel energy-aware and carbon-aware LLM system designs.

\section{Appraoch}

\subsection{Vidur Extension}

We extend Vidur with a power model that maps inference parameters, such as MFU, batch size, and parallelism, into GPU power draw and energy usage, which are then translated into carbon emissions using grid intensity data.

\vspace{-1.5em}
\subsubsection{GPU Power Modeling}

We model GPU power as a function of MFU, using a sublinear power-law to reflect the early saturation of power draw in memory-bound inference workloads~\cite{jiang2024megascale,trainy2024gpuutil}:

\begin{equation}
P(mfu) = P_{\text{idle}} + (P_{\text{max\_inst}} - P_{\text{idle}}) \cdot \left(\frac{mfu}{mfu_{\text{sat}}}\right)^\gamma
\label{eq:power_model}
\end{equation}

Here, $P_{\text{idle}}$ is the idle power draw, $P_{\text{max\_inst}}$ is the observed maximum under saturation, and $mfu_{\text{sat}}$ is the empirical saturation threshold. The exponent $\gamma < 1$ captures the diminishing increase in power as MFU rises.

Prior work suggests that GPU \textit{load}, captured by memory throughput, utilization, or hardware counters, correlates strongly with dynamic power draw during inference~\cite{patel2024characterizing,fu2024llmco2}. While our model does not use load directly, we adopt MFU as a practical proxy, as it is widely used to characterize LLM workloads and aligns with observed patterns: inference stages with lower MFU typically exhibit reduced power draw, especially during memory-bound operations like attention~\cite{chen2024efficient}. Sublinear power scaling across nodes further supports MFU’s utility in reflecting underutilization~\cite{fernandez2024hardware,trainy2024gpuutil}.

\vspace{-1.5em}
\subsubsection{Power Model Calibration}

We calibrated the power model using published specifications and benchmark data for NVIDIA A100 (SXM4), H100 (SXM5), and A40 (PCIe) GPUs. For the A100, we used 100~W (idle) and 400~W (peak), based on DGX server data~\cite{servethehome2023a100} and full-load benchmarks~\cite{horizoniq2023a100}. For the H100, we set 60~W and 700~W, reflecting its SXM5 configuration~\cite{megware2023h100}. The A40 was modeled with 30~W and 300~W, based on empirical data and sources~\cite{servethehome2023a40,nvidia2023a40}.

\vspace{-1.5em}
\subsubsection{Power Model Assumptions and Limitations}

Our MFU-based model estimates dynamic GPU power during LLM inference, capturing saturation from memory bottlenecks but omitting DVFS effects, cache behavior, memory overhead, and thermal throttling, which may reduce accuracy in real deployments. It is calibrated using public benchmarks~\cite{trainy2024gpuutil}, not direct telemetry, and has not been validated against profiling tools such as NVML. Idle and peak values reflect FP16/BF16 full-load behavior~\cite{servethehome2023a100,horizoniq2023a100,megware2023h100}, but may vary under power caps or shared workloads. The model assumes static per-iteration power and excludes host, interconnect, and cooling energy. It also relies on workload-specific MFU traces generated via Vidur; therefore, applying it to unseen workloads would require prior profiling, simulation, or estimation of MFU behavior.

\vspace{-1.5em}
\subsubsection{Energy Modeling and Carbon Accounting}

We estimate energy and carbon emissions by combining Vidur’s MFU-based output, logged at replica stage granularity, with GPU configuration and carbon intensity data.

\paragraph{Operational Energy.} For each batch stage $i$:
\begin{equation}
\text{MFU}_i = \frac{\text{FLOPs}_{\text{MLP}}(i) + \text{FLOPs}_{\text{Attention}}(i)}{\mathrm{DeviceFLOPs} \times t_i} \times 100, \quad
G = R \cdot \text{TP} \cdot \text{PP},
\end{equation}
\begin{equation}
H_i = \frac{\Delta t_i}{3600} \cdot G, \quad
E_{\text{op}} = \sum_{i=1}^{N} P(\text{MFU}_i) \cdot H_i \cdot \text{PUE}
\label{eq:energy_model}
\end{equation}
where $t_i$ is the replica stage duration and $P(\cdot)$ is the MFU–power model (Eq.~\ref{eq:power_model}).

\paragraph{Carbon Emissions.} Total emissions are:
\begin{equation}
C = E_{\text{op}} \cdot \text{CI} + H \cdot \phi_{\text{manuf}},
\label{eq:carbon_model}
\end{equation}
where CI is regional grid intensity (gCO$_2$/kWh) and $\phi_{\text{manuf}}$ is the per-GPU embodied carbon rate. Both static and time-varying CI are supported.

\subsection{Vidur–Vessim Integration}

We integrate Vidur with Vessim to simulate LLM inference energy use and carbon emissions under realistic grid conditions. This section outlines the modifications to Vidur, the Vessim setup, and the data pipeline linking them.

\vspace{-1.0em}
\subsubsection{Modifying Vidur for Vessim Compatibility}

To enable time-resolved simulation, we modified Vidur’s output to log MFU at the batch stage level instead of producing replica stage-wide average over the simulation duration. Each stage is recorded in a separate JSON file, capturing MFU distributions and execution times. Using Equation~\ref{eq:power_model}, each MFU entry is converted into instantaneous power. These values are aggregated into a unified DataFrame, indexed by timestamp and replica/stage IDs, which serves as input to Vessim.

\vspace{-1.0em}
\subsubsection{Vessim Setup}

Vessim~\cite{wiesner2024vessim} is a co-simulation framework for energy systems, integrating renewable generation, carbon intensity, and battery dynamics. In our setup:

\begin{itemize}[nosep]
    \item \textbf{Power Signal.} Vidur-generated power is resampled and wrapped in a \texttt{HistoricalSignal}.
    \item \textbf{Carbon Intensity.} WattTime grid intensity~\cite{watttime2023caisonorth} is used (gCO$_2$/kWh).
    \item \textbf{Solar Signal.} Solcast irradiance data~\cite{solcast2022global}, scaled by a configurable factor.
\end{itemize}

A \texttt{ClcBattery} object simulates storage with customizable state-of-charge constraints. Actors (e.g., \texttt{vidur\_power\_usage}, \texttt{solar}), controllers (e.g., \texttt{Monitor}, \texttt{CarbonLogger}), and an \texttt{Environment} component execute simulations at configurable resolution (default: 1 minute).

\vspace{-1.0em}
\subsubsection{Vidur–Vessim Data Pipeline}

Vidur’s MFU logs use variable-duration batch stages, while Vessim expects fixed-resolution time series. We bridge this mismatch via a three-step signal preprocessing pipeline:

\begin{enumerate}[leftmargin=*, itemsep=0.4em]
    \item \textbf{Timestamps.} Each batch stage is timestamped using Vidur’s internal clock.
    \item \textbf{Aggregation.} Convert power into 1-minute bins using a weighted average:
    \begin{equation}
    \bar{P} = \frac{\sum_{i=1}^{N} P_i \cdot \Delta t_i}{\sum_{i=1}^{N} \Delta t_i},
    \label{eq:avg_power}
    \end{equation}
    where \( P_i \) is instantaneous power and \( \Delta t_i \) is batch stage duration.
    \item \textbf{Export.} Save to CSV in Vessim’s load profile format.
\end{enumerate}

This produces a structured, time-aligned power profile for carbon-aware co-simulation. All pipeline parameters (e.g., battery size, solar scale, aggregation interval) are user-configurable via CLI or config files.

\vspace{-1.0em}
\subsubsection{Integration Assumptions and Limitations}

We balance fidelity and tractability via several simplifications. Environmental datasets (Solcast, WattTime) are resampled via cubic interpolation to match Vessim’s resolution, which may introduce artifacts. We also simulate non-seasonal alignment (e.g., applying June–July traces to winter workloads). Vidur is most accurate near 85\% of its max QPS~\cite{agrawal2024vidur} and requires careful calibration. We model only GPU-local energy and exclude CPU, memory, and cooling. These limitations constrain generalizability but allow detailed exploration of time-resolved dynamics. Future work could address telemetry-based calibration and improved seasonal matching.

\section{Results and Evaluation}

We structure our results in three parts: (1) experimental setup and parameterization, (2) modeling insights and controlled Vidur simulations, and (3) a case study demonstrating Vidur–Vessim integration.

\subsection{Experimental Setup}

\subsubsection{Power Model Parameters}

We parameterize the GPU power model (Eq.~\ref{eq:power_model}) for the NVIDIA A100 (80GB SXM4). Idle and peak power values are set to 100~W and 400~W, respectively, based on vendor specs and benchmark data. The MFU saturation threshold is set to $mfu_{\text{sat}} = 0.45$, and the scaling exponent to $\gamma = 0.7$, reflecting sublinear power scaling observed during inference\footnote{$\gamma = 0.7$ was heuristically chosen to reflect a plausible sublinear relationship between MFU and power in inference workloads. It was not fit to data due to the lack of real GPU power traces in our simulation, but future work will refine the model using empirical measurements.}.

\begin{figure}[t]
    \centering    \includegraphics[width=0.6\linewidth]{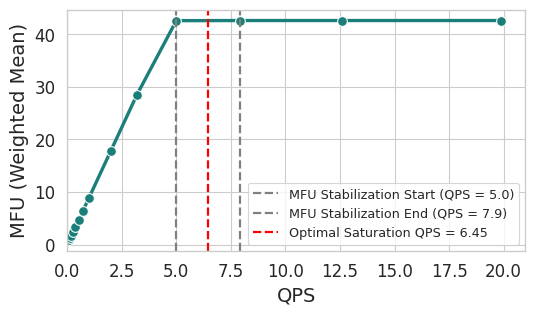}
    \vspace{-1em}
    \caption{\scriptsize
    Simulated QPS saturation for Meta-Llama-3-8B. MFU increases with QPS before plateauing at 5--7.9 QPS. The dashed line marks the saturation threshold near $mfu_{\text{sat}} = 0.45$. See Section~\ref{sec:exp4-qps-power}.
    }
    \label{fig:qps_sat}
\end{figure}

The threshold $mfu_{\text{sat}} = 0.45$ reflects both empirical observations from our simulations and patterns documented in industry analyses of transformer-based LLMs. Specifically, prior profiling efforts have shown that memory bandwidth limitations often cap MFU well below peak GPU compute utilization.  For instance, recent work by Trainy~\cite{trainy2024gpuutil} highlights that despite achieving 100\% GPU utilization, LLM workloads frequently plateau around 35\%–45\% MFU due to memory-bound bottlenecks such as softmax and MLP operations. Similarly, Figure~\ref{fig:qps_sat} shows MFU saturating near 0.45 under QPS values of 5–7.9. These consistent saturation patterns motivate our selection of 0.45 as the maximum effective utilization level in our power model.

\vspace{-1em}
\subsubsection{Simulation and Integration Parameters}

All experiments use a common set of Vidur parameters (Table~\ref{tab:side_by_side_configs}, Panel (a)), with request lengths drawn from a Zipfian distribution to reflect the power-law structure of language data~\cite{neumann2024alphazipf}. For the carbon-aware integration, we modify select Vidur parameters and introduce environment-specific Vessim settings, summarized in Panel (b) of Table~\ref{tab:side_by_side_configs}.

\begin{table}[t]
\centering
\caption{\scriptsize Simulation and integration parameters.}
\vspace{-0.5em}
\label{tab:side_by_side_configs}
\scriptsize
\renewcommand{\arraystretch}{0.9}
\begin{minipage}[t]{0.48\textwidth}
\centering
\textbf{(a) Default Vidur Configuration}\\[0.3em]
\begin{tabular}{|p{0.38\linewidth}|p{0.54\linewidth}|}
\hline
\textbf{Param.} & \textbf{Value} \\
\hline
Device & NVIDIA A100 (80GB) \\
Model & Meta-Llama-3-8B \\
TP / PP & 1 / 1 \\
Exec. Model & Random Forest ($k$=10) \\
Req. Length & Zipf \\
QPS & 6.45 (Poisson) \\
Scheduler & vLLM, RR \\
Batch Cap & 128 \\
Max Tokens & 4096 \\
Requests & 1,024 \\
PUE & 1.2 (CA)~\cite{google_datacenter_efficiency} \\
\hline
\end{tabular}
\end{minipage}
\hfill
\begin{minipage}[t]{0.48\textwidth}
\centering
\textbf{(b) Vidur--Vessim Integration}\\[0.3em]
\begin{tabular}{|p{0.38\linewidth}|p{0.54\linewidth}|}
\hline
\textbf{Param.} & \textbf{Value} \\
\hline
Model & Llama-2-7B-hf \\
Requests & 400,000 \\
Req. Length & Zipf ($\theta$=0.6, 1K–4K) \\
Prefill:Decode & 20.0 \\
QPS & 20 (Poisson) \\
Topology & NVLink (pairwise) \\
Location & CAISO-North \\
Solar Cap. & 600 W \\
Battery & 100 Wh, SoC: 80\% / 20\% \\
Carbon Thresh. & 100 / 200 gCO\textsubscript{2}/kWh \\
Interval & 1 min \\
\hline
\end{tabular}
\end{minipage}
\end{table}

\subsection{Controlled Simulation Experiments}\label{sec:controlled-sim-experiments}

We ran five Vidur simulations, each varying specific parameters from the default setup (Table~\ref{tab:side_by_side_configs}, Panel (a)), to isolate the effects of request volume, model architecture, and parallelism on GPU power and energy during LLM inference.

\begin{figure}[hbtp]
\centering
\includegraphics[width=0.9\linewidth]{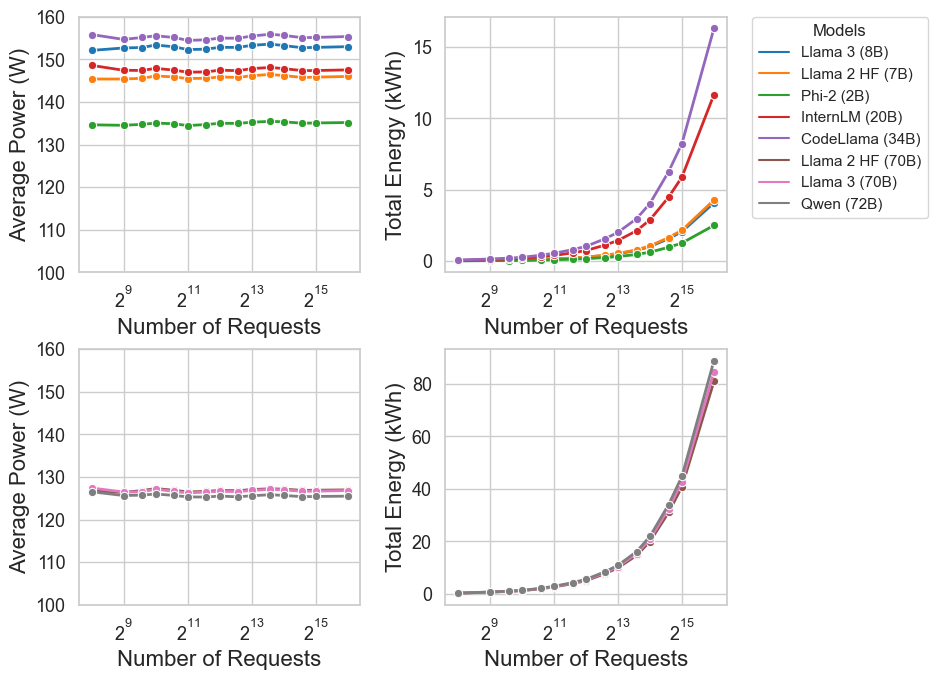}
\vspace{-0.5em}
\caption{\scriptsize
Average power draw and total energy usage as a function of request count. Top: models up to 34B. Bottom: 70B+ models. All runs use A100 GPUs with Zipfian request distributions.
}
\label{fig:experiment_1}
\bigskip
\bigskip
\bigskip
\includegraphics[width=0.9\linewidth]{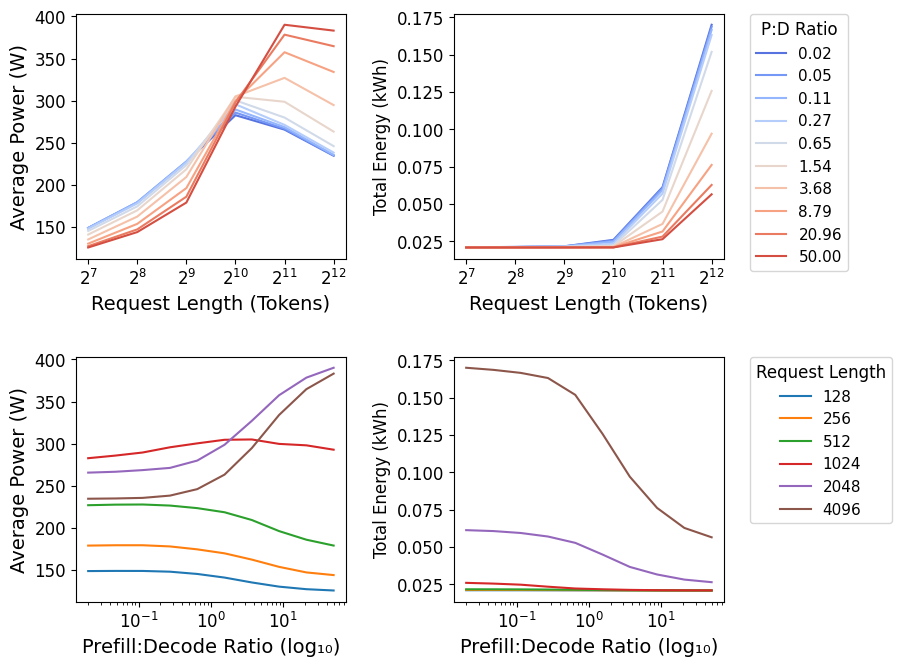}
\vspace{-0.5em}
\caption{\scriptsize
Impact of prefill-to-decode (P:D) ratio on power and energy usage. Panels (A) and (B): grouped by P:D ratio across varying request lengths. Panels (C) and (D): grouped by request length across varying P:D ratios.
}
\label{fig:experiment_2}
\end{figure}

\vspace{-1em}
\subsubsection{Experiment 1: Number of Requests vs. Power and Energy Usage}

We vary request counts from $2^8$ to $2^{16}$ across models ranging from 2.7B to 72B parameters. Models up to 34B use TP=1, PP=1; 70B+ models use TP=2, PP=2. As shown in Figure~\ref{fig:experiment_1}, average GPU power remains stable: 135–155~W for models $\leq$34B, and 125–127.5~W for 70B+ models. In contrast, total energy usage increases approximately linearly with request volume, despite appearing exponential due to the logarithmic x-axis scale. At $2^{16}$ requests, CodeLLaMA-34B reaches $\sim$16~kWh, while LLaMA-3-70B and Qwen-72B exceed 80~kWh.

\vspace{-1em}
\subsubsection{Experiment 2: Prefill-to-Decode Ratio vs. Power and Energy Usage}

We vary prefill-to-decode (P:D) ratios logarithmically from 50:1 to 1:50 across fixed request lengths (128–4096 tokens). As shown in Figure~\ref{fig:experiment_2}, at fixed P:D ratios, both power and energy generally increase with request length (Panels A–B). At fixed lengths, increasing the P:D ratio (i.e., more decode-heavy) leads to higher power and energy usage, especially for long requests; short requests show little change (Panels C–D).

\vspace{-1em}
\subsubsection{Experiment 3: Batch Size vs. Power and Energy Usage}

\begin{figure}[t]
\centering
\includegraphics[width=0.9\linewidth]{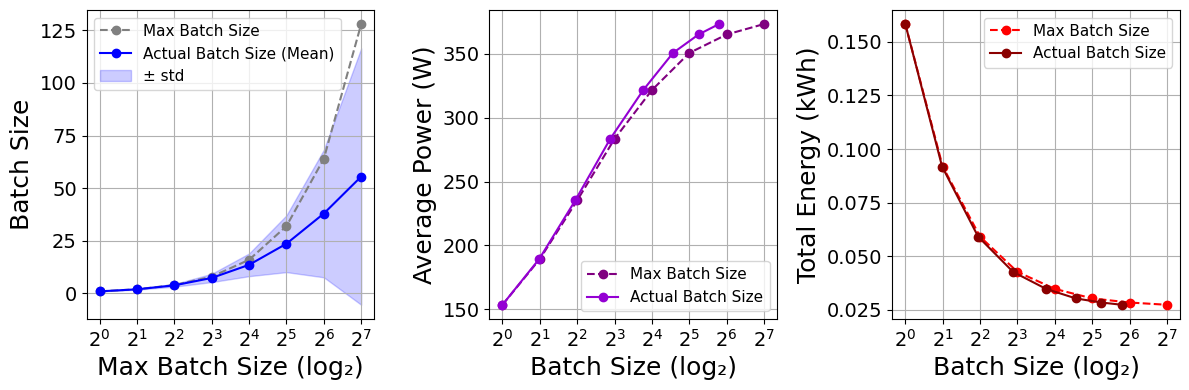}
\vspace{-0.5em}
\caption{\scriptsize
Effect of batch size cap on power and energy metrics. (A) Actual vs. configured batch size. (B) Average GPU power draw. (C) Total energy usage.
}
\label{fig:experiment_3}
\end{figure}

We vary the batch size cap from 1 to 128 to study its effect on GPU power and energy. As shown in Figure~\ref{fig:experiment_3}, actual batch size increases sublinearly with the cap, with high variance beyond 32 (Panel A). Average power rises with batch size, reaching a plateau above cap 64 (Panel B). Total energy consumption drops consistently with larger batches, with decreasing returns past cap 16 (Panel C).

\begin{figure}[t]
\centering
\includegraphics[width=0.8\linewidth]{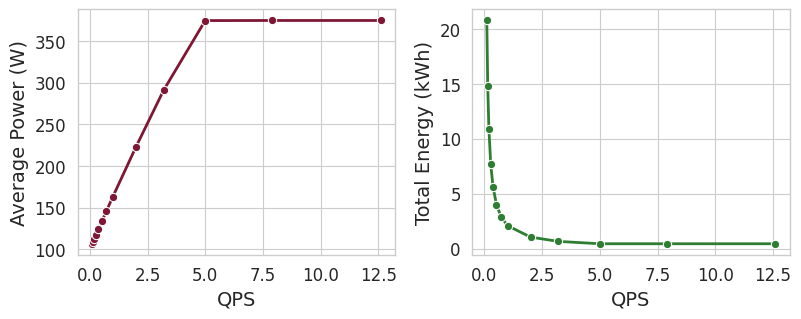}
\vspace{-0.5em}
\caption{\scriptsize
Effect of query throughput (QPS) on power and energy metrics. (A) Average GPU power draw across QPS values. (B) Total energy usage for a fixed workload of $2^{14}$ requests.
}
\label{fig:experiment_4}
\end{figure}

\vspace{-1em}
\subsubsection{Experiment 4: Query Throughput (QPS) vs. Power and Energy Usage}\label{sec:exp4-qps-power}

We vary QPS from 0.1 to 50 in a logarithmic interval while fixing the request count at $2^{14}$, but display the results only for QPS~$\leq$~12.6 to focus on the regime where power and energy trends stabilize (Figure~\ref{fig:experiment_4}). As shown in Panel~A, average GPU power increases with QPS and saturates near 360\,W beyond QPS~$\approx$~5. In Panel~B, total energy decreases with QPS due to shorter execution time and converges toward 0.5\,kWh beyond QPS~$\approx$~8.

\vspace{-1em}
\subsubsection{Experiment 5: Parallelism Configuration vs. Power and Energy Usage}

We vary tensor parallelism (TP) and pipeline parallelism (PP) in \{1, 2, 4\} using CodeLlama-34B on a 4$\times$A100 cluster with NVLink, resulting in 9 TP–PP configurations. Average GPU power range from 213.2\,W to 355.3\,W, peaking at TP=2, PP=1 and dropping with higher parallelism. Energy usage vary between 0.16–0.56\,kWh, with the most efficient setups (TP=2, PP=1 and TP=1, PP=2) balancing runtime and power draw. These results suggest that energy efficiency is more strongly influenced by reduced runtime than by minimizing power draw.

\subsection{Vidur--Vessim Integration for Carbon-Aware Inference}\label{sec:vidur-vessim-integration}

We simulate a deployment scenario by linking Vidur’s inference modeling with Vessim’s time-resolved grid simulation (parameters in Table~\ref{tab:side_by_side_configs}, Panel (b)). This co-simulation enables carbon-aware evaluation informed by Section~\ref{sec:controlled-sim-experiments}.

\begin{table}[t]
\centering
\caption{\scriptsize Summary of energy, battery, and emissions metrics from the Vidur--Vessim illustration.}
\vspace{-0.5em}
\label{tab:integration_summary}
\scriptsize
\renewcommand{\arraystretch}{0.95}
\begin{tabularx}{\textwidth}{|X|c||X|c|}
\hline
\textbf{Metric} & \textbf{Value} & \textbf{Metric} & \textbf{Value} \\
\hline
Total energy demand & 5.90~kWh & Avg. SoC & 47.2\% \\
Solar generation & 4.15~kWh & Time $<$ 50\% SoC & 15.7~h \\
Grid consumption & 1.81~kWh & Time $>$ 80\% SoC & 6.7~h \\
Renewable share & 70.3\% & Charging duration & 21.2\% \\
Grid dependency & 30.7\% & Discharging duration & 14.0\% \\
Total emissions & 2.47~kgCO\textsubscript{2} & Idle time & 64.8\% \\
Offset by solar & 1.71~kgCO\textsubscript{2} & Carbon offset & 69.2\% \\
Net footprint & 759.2~gCO\textsubscript{2} & Avg. carbon intensity & 418.2~gCO\textsubscript{2}/kWh \\
Time in high-CI hours & 24.8~h & Battery Full Cycles & 0.8 \\
\hline
\end{tabularx}
\end{table}

\begin{figure}[t]
\centering
\vspace{0.5em}
\includegraphics[width=0.7\linewidth]{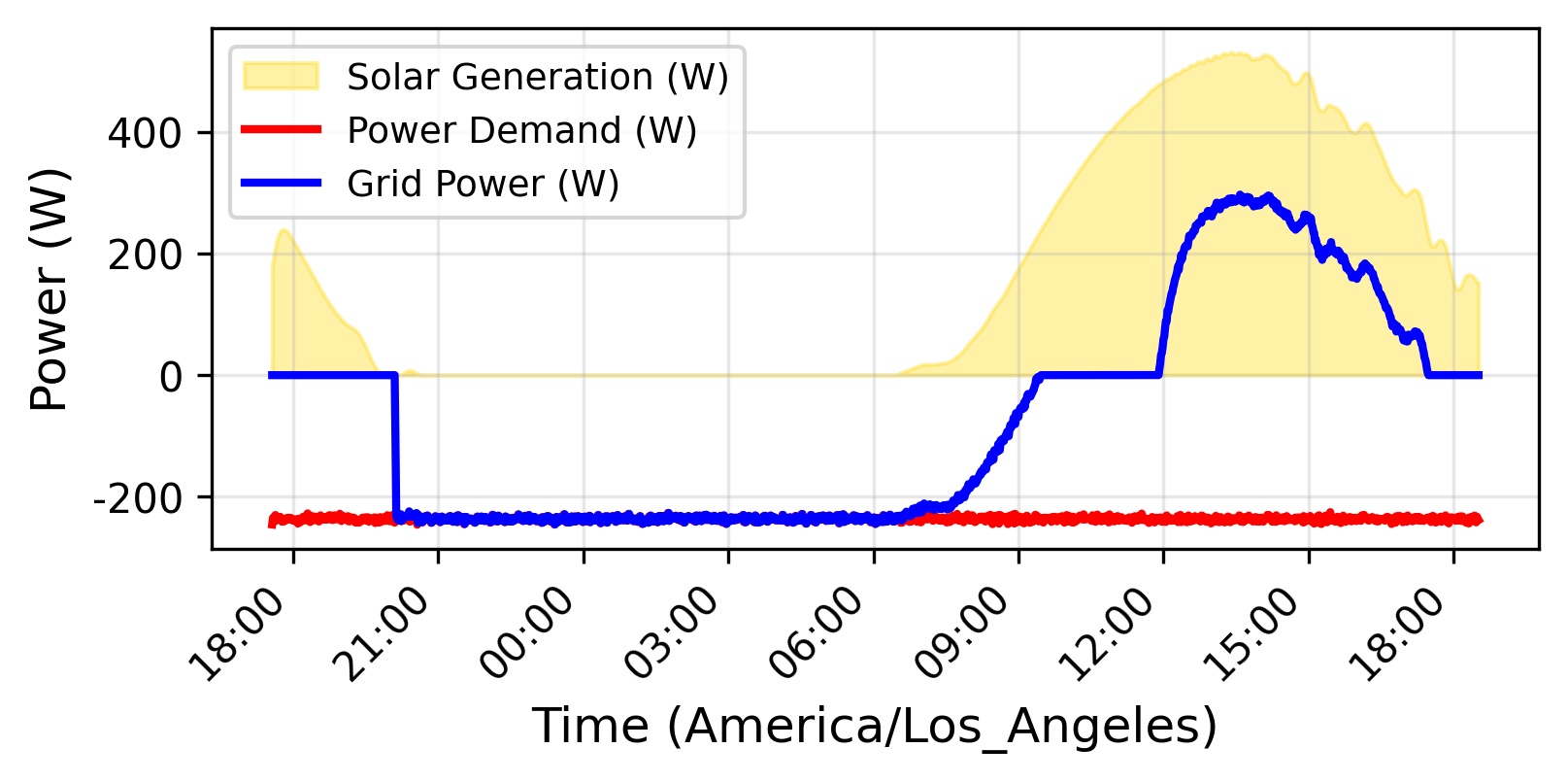}
\vspace{-1em}
\caption{\scriptsize
Time-resolved power flow for the LLM inference workload. Solar generation (yellow) peaks midday and partially offsets the workload's power demand (red). Net grid power (blue) reflects residual demand not met by solar, transitioning from negative (grid export) to positive (grid draw) depending on solar availability and battery state.
}
\label{fig:power_flow}
\end{figure}

\begin{figure}[t]
\centering
\begin{minipage}[t]{0.48\linewidth}
    \centering
    \includegraphics[width=\linewidth]{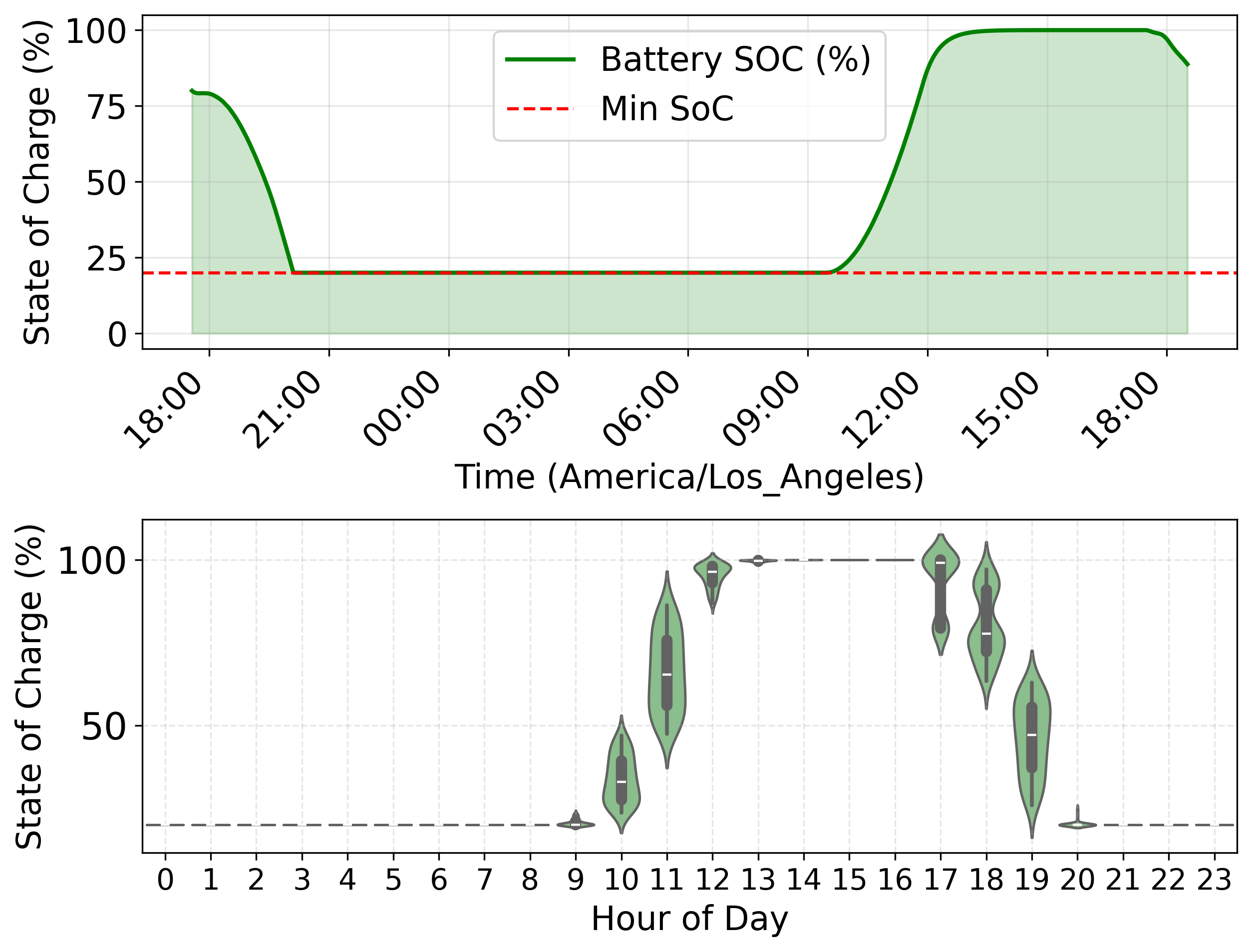}
    \label{fig:battery_soc}
\end{minipage}
\hfill
\begin{minipage}[t]{0.48\linewidth}
    \centering
    \includegraphics[width=\linewidth]{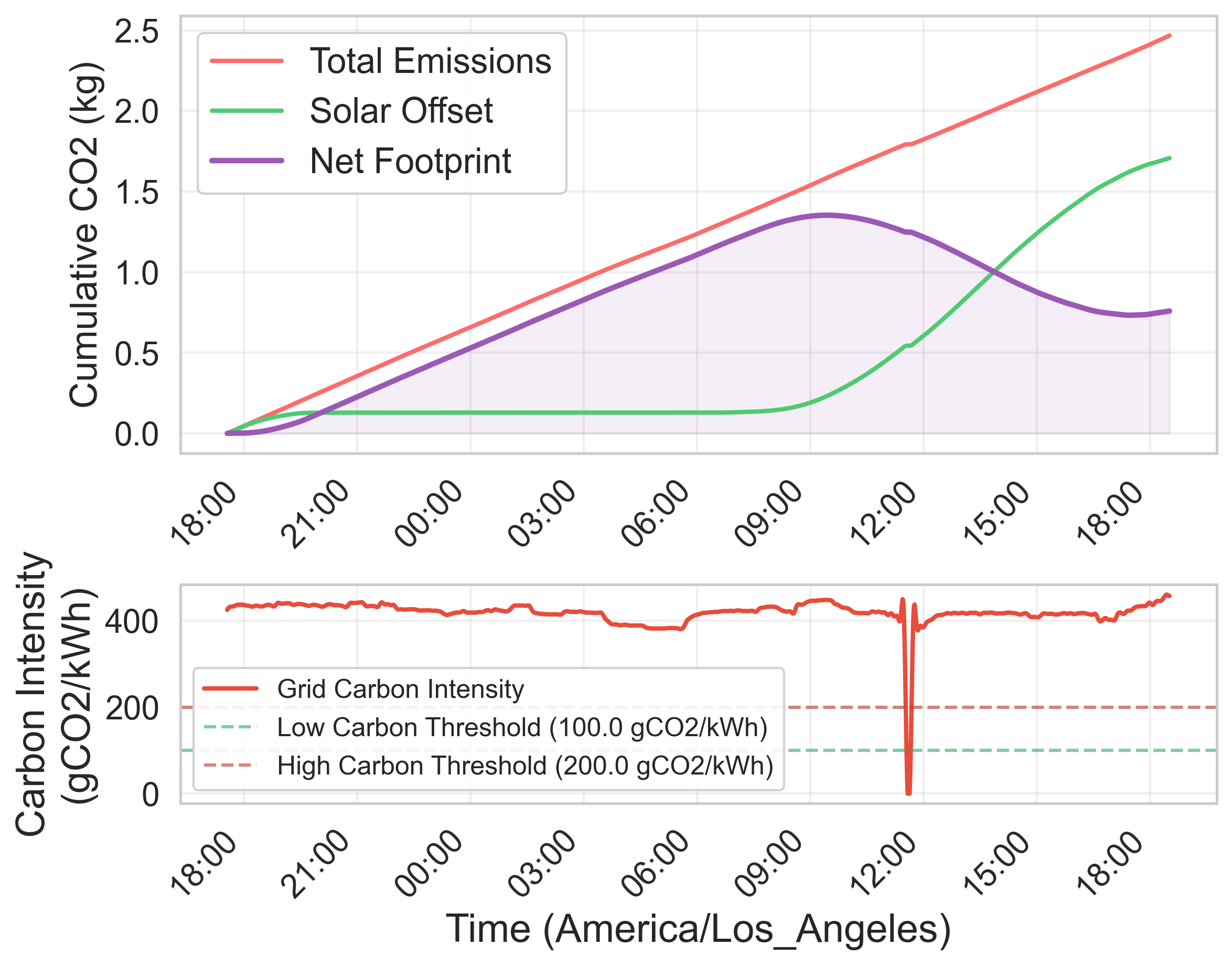}
    \label{fig:carbon_analysis}
\end{minipage}
\vspace{-1em}
\caption{\scriptsize
Battery and emissions behavior during Vidur--Vessim co-simulation. Left: battery state of charge (SoC) trace and hourly SoC distribution. Right: cumulative carbon emissions, solar offset, net footprint, and grid carbon intensity trace over time with high and low carbon intensity thresholds.
}
\label{fig:battery_and_carbon}
\end{figure}

The workload drew 5.90~kWh in total, 70.3\% of which was supplied by solar (4.15~kWh) and the rest by grid energy (1.81~kWh) (Table~\ref{tab:integration_summary}, Fig.~\ref{fig:power_flow}). Battery usage was minimal: 0.8 cycles, 47.2\% average state of charge (SoC), and 64.8\% idle time. Total emissions reached 2.47~kgCO\textsubscript{2}, of which 69.2\% were offset by solar. The system operated under high-carbon-intensity conditions ($>$200~gCO\textsubscript{2}/kWh) for 24.8 hours, with an average CI of 418.2~gCO\textsubscript{2}/kWh (Fig.~\ref{fig:battery_and_carbon}).

This case study highlights that while the majority of energy demand was met by solar, battery cycling was limited and much of the workload ran during high-carbon hours. These dynamics emphasize the value of time-resolved co-simulation in understanding the interaction between renewable supply, storage behavior, and emissions across realistic workload traces.

\section{Discussion and Future Directions}

Our simulations surface key tradeoffs between energy efficiency and performance in distributed LLM inference. Increasing query throughput (QPS) reduces total energy by shortening execution time, but also pushes GPUs into high-power regimes. Larger batch sizes improve efficiency up to a point, after which variance and latency begin to erode gains.

Parallelism configurations add further complexity. More GPUs (e.g., TP=4, PP=4) do not guarantee better efficiency, as our most energy-efficient setting was a moderate configuration (TP=2, PP=1), balancing saturation and interconnect overhead.
At the system level, our Vidur–Vessim integration reveals that renewable availability alone is insufficient to minimize emissions. Even with 70\% solar coverage, fixed scheduling and idle battery behavior led to high-carbon execution. This highlights the importance of real-time, grid-aware adaptation.

We envision a fully coupled co-simulation loop: Vidur dynamically adjusts inference parameters in response to Vessim’s evolving grid signals, while Vessim adapts datacenter behavior to simulated workloads. This bidirectional feedback enables more realistic modeling of datacenter decision-making under environmental constraints.

Our framework also extends naturally to multi-region routing. With grid intensity data and modeled interconnect costs, it can simulate when and where to shift inference jobs to lower-carbon regions, raising new optimization questions at the intersection of performance, compliance, and sustainability.

Finally, our work sets the stage for exploring policy tradeoffs, such as deploying smaller models in high-CI regions versus larger ones during renewable peaks. While direct validation against empirical power data is outside the present scope, we recognize this as a key limitation. Future work will incorporate telemetry-based calibration, expanded modeling of memory and thermal dynamics, and real-time simulation control. The core contribution of this study is the development of a flexible, extensible framework for analyzing carbon-aware inference strategies, which can readily integrate empirical data as it becomes available.

\section{Conclusion}

We present a simulation framework that links high-fidelity inference modeling with time-resolved energy use and carbon emissions simulation to evaluate sustainable LLM deployment strategies. By integrating Vidur and Vessim, we enable detailed analysis of how workload parameters, like batching, throughput, and parallelism, interact with both hardware behavior and dynamic grid conditions.

Our findings show that configurations optimized for performance, such as those with the highest query throughput (QPS) or hardware utilization (MFU), do not always yield the lowest energy or carbon usage. Our framework surfaces these tradeoffs and enables dynamic, carbon-aware adaptation to grid signals and renewable availability. As LLM deployments scale across distributed and concurrent systems, tools like ours will be essential for energy efficiency and performance optimization under realistic, environmentally constrained conditions.
%
%
%
\bibliographystyle{splncs04}
\bibliography{references}

\end{document}